\documentclass[aps,prl,showpacs,twocolumn]{revtex4-1}
\usepackage{epsfig}
\usepackage{color}
\usepackage{hyperref}

\definecolor{red}{rgb}{1,0,0}
\definecolor{blue}{rgb}{0,0,1}
\definecolor{black}{rgb}{0,0,0}

\newcommand{\be}{\begin{equation}}
\newcommand{\ee}{\end{equation}}
\newcommand{\ba}{\begin{eqnarray}}
\newcommand{\ea}{\end{eqnarray}}

\newcommand{\figref}{Fig.~\ref}

\newcommand{\Tr}[1] {\textrm{Tr}\left[#1\right]}
\newcommand{\SYM}[1] {\left[#1\right]^{\rm S}}
\newcommand{\ASYM}[1] {\left[#1\right]^{\rm A}}
\newcommand{\ST}[1] {\left[#1\right]^{\rm ST}}

\newcommand{\tsrt}[1] {\tilde{\mathbf{#1}}}
\newcommand{\tsr}[1] {\mathbf{#1}}
\newcommand{\vtr}[1] {\mathbf{#1}}
\newcommand{\uvtr}[1] {\hat{\mathbf{#1}}}
\newcommand{\gdot}[0] {\dot{\gamma}}

\usepackage{amsmath}

\newlength{\arrow}
\begin{document}
\title{Active Viscoelastic Matter: from Bacterial Drag Reduction to Turbulent Solids}

\author{E. J. Hemingway$^1$, A. Maitra$^2$, S. Banerjee$^{3,4}$, M. C. Marchetti$^3$, S. Ramaswamy $^{5,2}$, S. M. Fielding$^1$  and M. E. Cates$^6$  }

\affiliation{$^1$Department of Physics, Durham University, Science Laboratories, South Road, Durham DH1 3LE, United Kingdom}
\affiliation{$^2$CCMT, Department of Physics, Indian Institute of Science, Bangalore 560 012, India}
\affiliation{$^3$Physics Department and Syracuse Biomaterials Institute, Syracuse University, Syracuse, New York 13244, USA}
\affiliation{$^4$James Franck Institute, University of Chicago, Chicago, IL 60637, USA}
\affiliation{$^5$TIFR Centre for Interdisciplinary Sciences, 21 Brundavan Colony, Osman Sagar Road, Narsingi, Hyderabad 500 075 India}
\affiliation{$^6$SUPA, School of Physics and Astronomy, University of Edinburgh, James Clerk Maxwell Building, Peter Guthrie Tait Road, Edinburgh EH9 3FD, United Kingdom }
\date{\today}

\begin{abstract}
  A paradigm for internally driven matter is the active nematic liquid crystal, whereby the equations of a conventional nematic are supplemented by a minimal active stress that violates time-reversal symmetry. In practice, active fluids may have not only liquid-crystalline but also viscoelastic polymer degrees of freedom. Here we explore the resulting interplay by coupling an active nematic to a minimal model of polymer rheology. We find that adding polymer can greatly increase the complexity of spontaneous flow, but can also have calming effects, thereby increasing the net throughput of spontaneous flow along a pipe (a `drag-reduction' effect). Remarkably, active turbulence can also arise after switching on activity in a sufficiently soft elastomeric {\em solid}.
\end{abstract}

\maketitle

Active materials include bacterial swarms in a fluid, the cytoskeleton of living cells, and `cell extracts' containing just filaments, molecular motors, and a fuel supply \cite{RMP,Cisneros,Actomyosin,Dogic}. Such materials are interesting because of their direct biophysical significance, and as representatives of a wider class of systems in which deviations from thermal equilibrium are not created by initial or boundary conditions (a temperature quench, or motion of walls in a shear cell) but arise microscopically in the dynamics of each particle. By continually converting chemical energy into motion, active matter violates time-reversal symmetry, suspending the normal rules of thermal equilibrium dynamics (until the fuel runs out), causing strongly non-equilibrium features such as spontaneous flow. This flow may remain steady and laminar at the scale of the system; may show limit cycles at that scale or below; or may show spatiotemporal chaos.  Since it resembles the inertial turbulence of a passive Newtonian fluid, the latter outcome is commonly called `bacterial' (or `active') turbulence \cite{Cisneros,MarenduPRE,YeomansPNAS,CristinaDefect,CristinaDefectPreprint,Thampi}. The mechanism is quite distinct, however, stemming from
a balance between active stress and orientational relaxation, rather than between inertia and viscosity as in conventional turbulence.

The phenomenology of activity-driven spontaneous flow can be understood, to a remarkable extent, using conceptually simple continuum models \cite{RMP,SRreview,Simha,Joanny}. These start from the hydrodynamic equations of a passive fluid of rod-like objects with either polar \cite{Joanny} or nematic \cite{Simha} local order, the latter characterized by a tensor order parameter ${\bf Q}({\bf r})$ \cite{Simha}. To the passive equations for such a liquid crystal (LC) \cite{BerisEdwards} are then added leading-order violations of time-reversal symmetry; after renormalization of passive parameters and allowing for fluid incompressibility, what remains is a bulk stress $\tsr{\Sigma}_A = -\zeta {\bf Q}$ where $\zeta$, an activity parameter, is positive for extensile systems, negative for contractile. In extensile materials each rodlike particle pulls fluid inwards equatorially and emits it symmetrically from the poles, with the reverse for the contractile case. Even without accurate knowledge of $\zeta$, this approach makes robust predictions. For example, extensile and contractile systems become separately unstable toward spontaneous flow states at critical activity levels that are system-size dependent, and vanish for bulk samples. Numerical solution of the active nematic equations \cite{MarenduPRE,CristinaDefect,CristinaDefectPreprint,Thampi} show spontaneous flows resembling experiments on bacterial swarms \cite{Cisneros} and on microtubule-based cell extracts \cite{Dogic}. Both of these are extensile nematics, and we restrict ourselves to this case below \cite{AsterNote}.

Active nematogenic fluids are often referred to as `active gels' \cite{Joanny,MarenduPRE}. But although all LCs are somewhat viscoelastic (due to slow defect motion) these models assume fast local relaxations and mostly do not address gels in a conventional sense \cite{Callan}. Certainly they do not capture the diversity of viscoelastic behavior that one expects in sub-cellular active matter containing long-chain flexible polymers, or other cytoplasmic components, with long (possibly divergent) intrinsic relaxation times. These slow relaxations should couple to the orientational order, strongly modifying the effects of activity.  Polymers could also play a strong role in modifying diffusion \cite{Contrary} and active flows at supra-cellular level: they are present in mucus, saliva, and other viscoelastic fluids in which swarms of motile bacteria reside. Notably, many bacteria excrete their own polymers \cite{Exo-polymer}, suggesting an advantage in controlling the viscoelasticity of their surroundings.

In this Letter, therefore, we present a model that addresses the interplay between active LC and polymers \cite{Callan}. We sketch its derivation (which requires care) and give examples of its rich dynamics (which will be explored further in \cite{EwanLong}). Highlights include: an exotic form of `drag reduction' by polymers for active (non-inertial) turbulence; spontaneous flows with slow polymer-driven oscillations; and transient active turbulence within a material that is ultimately a solid.

{\it Equations of Motion:} The symmetric and antisymmetric parts of
the centre-of-mass velocity-gradient tensor
$\left(\nabla \vtr{v}\right)_{ij} \equiv \partial_i v_j$ are denoted
$\tsr{D}$ and $\tsr{\Omega}$ \cite{FootIndex}.  For other tensors the
symmetric, antisymmetric, and traceless parts carry superscripts S,A
and T. Conformation tensors for the polymer and LC are denoted
$\tsr{C}$ and $\tsr{Q}$, where $\tsr{Q}$ is traceless.  The polymeric tensor is $\tsr{C}=\langle{\bf rr}\rangle$, where ${\bf r}$ is
the end-to-end vector of a chain (or subchain, depending on the level of description). We
introduce a free energy density $f = f_Q(\tsr{Q},\nabla\tsr{Q}) +
f_C(\tsr{C})+ f_{QC}(\tsr{Q},\tsr{C})$ where $f_{Q}$ and $f_C$ are
standard forms for active nematics \cite{BerisEdwards} and dumb-bell
polymers \cite{Milner} respectively, as detailed in \cite{SI}. The
lowest order passive coupling is \be f_{QC} = \kappa
\Tr{\tsr{C}-\tsr{I}}\Tr{\tsr{Q}^2} + 2 \chi \Tr{\tsr{C}\tsr{Q}}.  \ee
Both terms vanish for undeformed polymers ($\tsr{C}=\tsr{I}$).

From the free energy $F = \int f dV$ we next derive the nematic molecular field $\tsr{H} \equiv -(\delta F/\delta \tsr{Q})^{\rm ST}$ as
\ba
\tsr{H} &= -G_Q\left[\left(1 - \frac{\gamma}{3}\right)\tsr{Q} - \gamma \tsr{Q}^2 + \gamma \tsr{Q}^3 \right] - G_Q \gamma \frac{\tsr{I}}{3}\Tr{\tsr{Q}^2} \nonumber\\ &+ K \nabla^2 \tsr{Q}
-2 \kappa \Tr{\tsr{C}-\tsr{I}}\tsr{Q} - 2\chi\tsr{C}^{\rm T}. \label{H}
\ea
Here $G_Q$ is a bulk free energy density scale set by $f_Q$; $K$ is the nematic elastic constant; $\gamma$ a control parameter for the nematic transition; and $G_C$ the polymer elastic modulus. (See details in \cite{SI}.) The corresponding molecular field for polymer conformations is simpler: $\tsr{B} \equiv -(\delta F/\delta \tsr{C}) = -G_C(\tsr{I}-\tsr{C}^{-1})/2 - \kappa \tsr I \Tr{\tsr{Q}^2}-2\chi\tsr{Q}$.

The most general equations of motion then involve at least four separate 4th-rank tensors describing how $\tsr{Q}$ and $\tsr{C}$ respond to these molecular fields, and to imposed velocity gradients. For simplicity we choose the response tensors of the Beris-Edwards LC theory and the Johnson-Segalman (JS) polymer model respectively \cite{BerisEdwards}. We then allow for conformational diffusion in the polymer sector \cite{StressDiffusion} which adds a gradient term in $\tsr{C}$ of kinetic origin \cite{SI}. The result is a minimally coupled model of the passive $\tsr{C}+\tsr{Q}$ dynamics that reduces to well-established models when either order parameter is suppressed.

To the coupled passive model we finally add a minimal set of active terms \cite{Simha}. In principle one can add all terms that violate time reversal symmetry arising at zeroth order in gradients and first order in either $\tsr{Q}$ or $\tsr{C}-\tsr{I}$; these are given in \cite{SI}. Here we suppose for simplicity that the polymers are not themselves active, and respond to nematic activity only through fluid advection. This captures the effect of adding polymer to (say) a cell extract; alternatively this could describe the collective dynamics of bacterial suspensions in mucus. (In contrast, one could build a system of polymers directly from active elements \cite{RjoyPol}.) There remain two active terms linear in $\tsr{Q}$; one can be absorbed into $f_Q$, and the other is the familiar active deviatoric stress $\tsr{\Sigma}_A = -\zeta\tsr{Q}$ \cite{Simha}.

The resulting equations of motion for $\tsr{Q}$ and $\tsr{C}$ are:
\ba
&\left(\partial_t + \vtr{v}.\nabla\right)\tsr{Q} = \tsr{Q}\tsr{\Omega} - \tsr{\Omega}\tsr{Q} + \frac{2\xi}{3} \tsr{D} + 2 \xi \ST{\tsr{Q}\tsr{D}} \nonumber \\
&- 2\xi \tsr{Q}\Tr{\tsr{Q}\tsr{D}} + \tau_Q^{-1} \tsr{H}/G_Q,%
\ea
\ba
&\left(\partial_t + \vtr{v}.\nabla\right)\tsr{C} = \tsr{C}\tsr{\Omega} - \tsr{\Omega} \tsr{C} + 2 a \SYM{\tsr{C}\tsr{D}} \nonumber \\
&+ \tau_C^{-1}(2\SYM{\tsr{B}\tsr{C}}/G_C+\ell^2_C\nabla^2 \tsr{C}).\label{EM2}
\ea
Here $\xi$ is the flow-alignment parameter of the nematic \cite{Stark} and $a$ is the slip parameter of the JS model. Each controls the relative tendency of molecules to align with streamlines and rotate with local vorticity. Parameters $\tau_Q,\tau_C$ are intrinsic relaxation times for nematic and polymer, while $\ell_C$ governs diffusion in the JS sector \cite{StressDiffusion}.

The incompressible fluid velocity $\vtr{v}$ obeys the Navier Stokes equation $ \rho \left(\partial_t + v_\beta\partial_\beta\right)v_\alpha = \partial_\beta\left(\Sigma_{\alpha\beta}\right)$ whose stress $\tsr{\Sigma} = -P\tsr{I} + 2\eta\tsr{D}+  \tsr{\Sigma}_A+\tsr{\Sigma}_{Q} + \tsr{\Sigma}_C $ combines an isotropic pressure $P$, a contribution from a Newtonian solvent of viscosity $\eta$, and active stress  $\tsr{\Sigma}_A$ with two reactive stresses \cite{Colon}
\ba
\tsr{\Sigma}_Q &=& - K(\nabla \tsr{Q}):(\nabla\tsr{Q}) + 2 \ASYM{\tsr{Q}\tsr{H}}  \nonumber \\
&-& \frac{2 \xi}{3} \tsr{H} - 2 \xi \ST{\tsr{Q}\tsr{H}}+ 2 \xi \tsr{Q} \Tr{\tsr{Q}\tsr{H}}, \label{QReactive}\\
\tsr{\Sigma}_C &=& -2a\SYM{\tsr{C}\tsr{B}} + 2 \ASYM{\tsr{C}\tsr{B}}.\label{CReactive}
\ea
Crucially, $\xi$ and $a$ must appear as shown in the reactive stresses to recover a correct passive limit \cite{BerisEdwards}. In the pure JS case, {\em but not in general}, one can absorb the factor $a$ in \eqref{CReactive} into $G_C$, restoring consistency to the classical JS model, which sets $\tsr{\Sigma}_C = -2\tsr{B}\tsr{C}$ for all $a$ \cite{BerisEdwards,Olmsted}. A less careful marriage of JS with active nematic theory would thus have set $a=1$ in \eqref{CReactive} but not \eqref{EM2}, violating thermodynamic principles \cite{Ottinger} and giving incorrect physics.

{\it Parameter Choices:} We choose $\xi$ and $a$ within the flow-aligning and outwith the shear-banding ranges of their respective models, to avoid tumbling and banding instabilities of the passive model in flow. We neglect inertia ($\rho = 0$), and choose units where $G_Q = \tau_Q = L_y = 1$, with $L_y$ the width of the sample, a 2D simulation box of $L_x\times L_y = 4\times 1$.  We choose periodic boundary conditions in $x$, with no-slip (of $\vtr{v}$) and no-gradient (of $\tsr{Q}$ or $\tsr{C}$) at the sample walls ($y=0,L_y$).
Default values for numerics are $\xi = 0.7,\eta = 0.567$ and $\gamma = 3$ (directly comparable with Ref. \cite{MarenduPRE} for the polymer-free case); we set $a=1$. We vary $\tau_C$ over several decades $10^{-2}\le\tau_C\le 10^6$ at fixed polymer viscosity $\eta_C\equiv \tau_CG_C = 1$, allowing fast or slow relaxation while retaining comparability of $\tsr{\Sigma}_{Q,C}$. We define $\ell_Q = (K/G_Q)^{1/2}$, the Frank length for nematic distortions, and vary this in the range $0.002 \le \ell_Q/L_y \le 0.025$ (comparable to other studies~\cite{MarenduPRE,YeomansPNAS}), and then set $\ell_C^2/\tau_C = \ell_Q^2/\tau_Q$ to equate the diffusivities of ${\bf Q}$ and ${\bf C}$. Using careful numerics we are able to address several decades of activity level $10^{-4}\le\zeta\le 6$.
Finally, most of our work addresses the simplest case where the coupling of $\tsr{Q}$ and $\tsr{C}$ is purely kinematic: {\em i.e.,} $\kappa = \chi = 0$. In this limit, interaction between polymer and $\tsr{Q}$ is indirect, mediated only via the background fluid velocity $\vtr{v}$. However we also present some results for nonzero $\chi$, as arises in passive nematic elastomers \cite{Warner}.

\begin{figure}
  \includegraphics[width=0.48\textwidth]{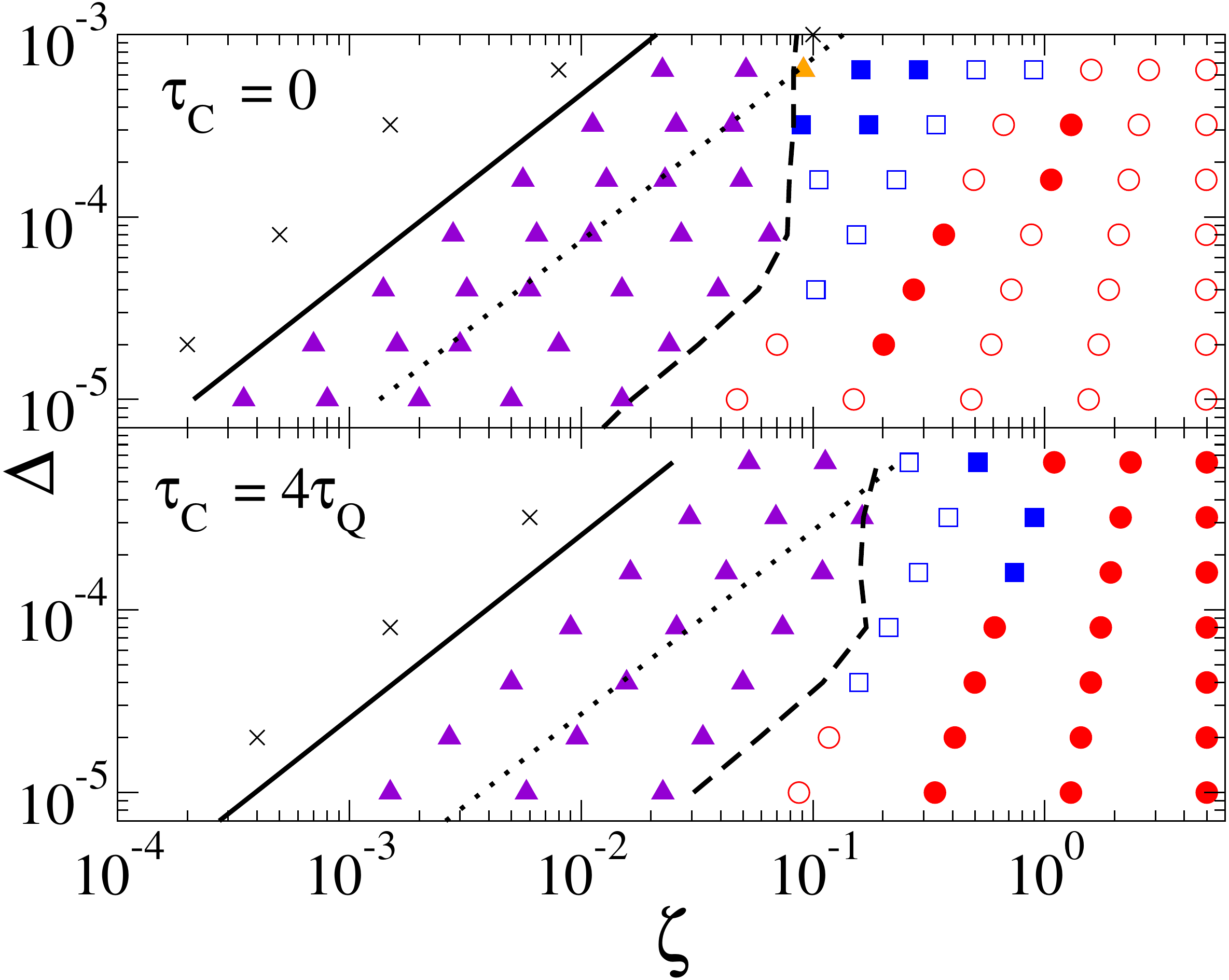}
  \\
  \vskip4pt
  \includegraphics[width=0.48\textwidth]{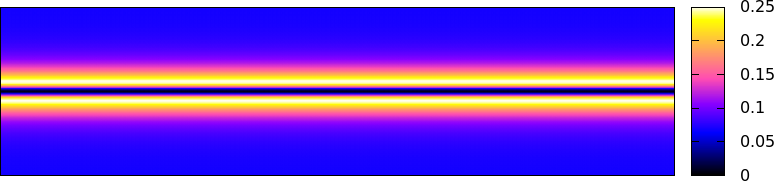}
  \\
  \vskip4pt
  \includegraphics[width=0.48\textwidth]{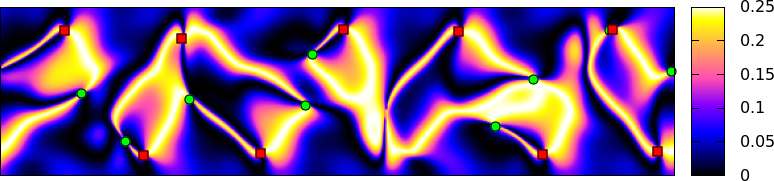}
  \\
  \vskip4pt
  \includegraphics[width=0.48\textwidth]{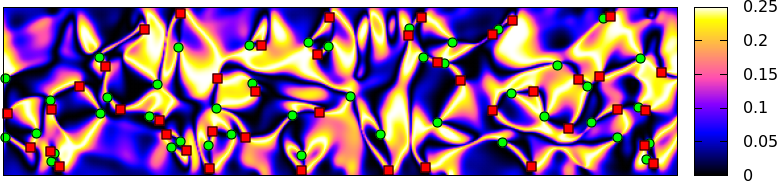}
  \caption{State diagrams without (upper) and with (lower) polymer of relaxation time $\tau_C=4\tau_Q$. Initial condition: director $\vtr{n}$ ({\em i.e.,} major axis of $\tsr{Q}$) uniformly along $y$. Symbols: $\times$: quiescent; squares: oscillatory; triangles: steady banded flow (cf \cite{MarenduPRE,FieldingPRL}), circles: unsteady/chaotic. Filled symbols denote states with a significant net throughput (along the periodic direction $x$) . Lines show (solid) the 1D instability (bending mode) of the specified initial condition; (dotted) that of the splay mode for initial condition with $\tsr{Q}$ along $x$, and (dashed) the observed crossover line $\zeta_c^{\rm bend 2D}$  beyond which the phase diagram becomes independent of which of these initial states was chosen.
    The bottom three panels show states, all with net throughput, from the $\tau_C = 4$ phase diagram above: banded ($\zeta=0.023$, $\Delta = 10^{-5}$), oscillatory ($\zeta = 0.741$, $\Delta = 1.6 \times 10^{-4}$) and chaotic ($\zeta = 1.75$, $\Delta = 8 \times 10^{-5}$); colour scale indicates $\left(n_x n_y\right)^2$. Defects of topological charge $\pm 1/2$ are identified by green dots (+) and red squares (-).
  \label{fig:PDs}}
\end{figure}

{\it Results:}
First, with kinematic coupling only, we ask whether addition of polymer can suppress the intrinsic instability of active nematics towards bulk flow. Generalizing previous results \cite{Joanny,EdwardsYeomans,MarenduPRE,Giomi}, a linear stability analysis (detailed in Ref. \cite{SI}) allowing 1D perturbations of wavevector $\vtr{k}$ about the quiescent nematic base state gives a critical activity level (for $\gamma = 3$)
\be
\zeta_c = \frac{12k^2\ell_Q^2}{\Lambda\tau_Q}\left(\eta + \frac{\Lambda^2
  G_Q\tau_Q}{72} +
\frac{a^2\eta_C}{1+k^2\ell^2_C}\right),\label{threshold}
\ee
where $\Lambda = 5\xi\pm3$ for $\vtr{k}$ perpendicular $(-)$ or parallel $(+)$ to the major axis of $\tsr{Q}$.
Thus $\zeta_c$ always vanishes in bulk (as $k\to 0$), while the final term shows a stabilizing effect of polymer in finite systems. This effect is viscous and not viscoelastic in character, since at threshold, the time-scale for growth diverges, with $\tau_C$ then infinitely fast in comparison.  This analysis, which we have confirmed numerically (\figref{fig:PDs}), contrasts with Ref.\cite{Contrary} which reports polymer-induced bulk stabilization for a related but distinct active model (with no inherent nematic tendency).

\figref{fig:PDs} shows phase diagrams on the $\zeta,\Delta$ plane, where $\Delta \equiv (\ell_Q/L_y)^2$ represents the stabilizing effect of small sample sizes. Varying $\tau_C$ at fixed $\eta_C=1$ reveals a very interesting effect of strictly viscoelastic origin. Among states showing active turbulence, adding polymer significantly extends the parameter range in which macroscopic symmetry is broken (filled symbols in~\figref{fig:PDs}), as judged by a criterion (see \cite{SI}) of significant net throughput of fluid along the (periodic) $x$ direction. Thus adding polymer to (say) a fluid showing bacterial turbulence should effectively `reduce drag' by enhancing throughput at fixed (active) stress -- as it does for pressure-driven turbulent pipe flow in a passive fluid \cite{DragReduct}.
The polymer calms the short scale structure of the active flow, decreasing the nematic defect density and increasing the flow correlation length towards the system size, thereby favoring restoration of a more organized flow state.

\begin{figure}
  \includegraphics[width=0.48\textwidth]{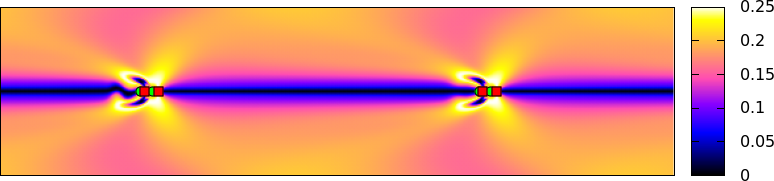}
  \vskip4pt
  \includegraphics[width=0.48\textwidth]{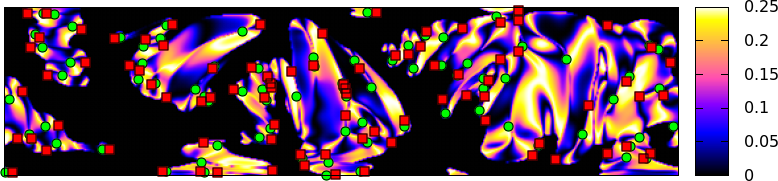}
  \vskip4pt
  \includegraphics[width=0.48\textwidth]{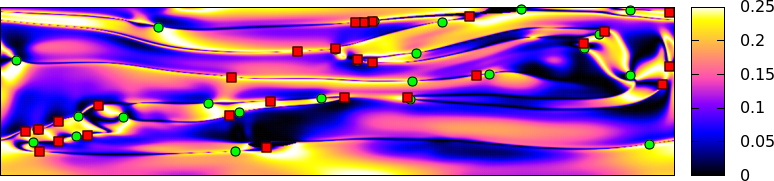}
  \caption{Three spontaneous flow states seen with added polymer, all with $\tau_C = 10$.
    Upper frame: a pair of defects traveling along the interface of a shear-banded state ($\zeta = 3.2$, $\Delta = 10^{-4}, \chi = 0.002$).
    Middle frame: coexistence of `bubbling' active domains and regions where the director is out of plane (black) ($\zeta = 6$, $\Delta = 10^{-4}$, $\chi = 0.004$).
  Lower frame: an exotic oscillatory state which coherently `shuffles' left and right on timescale $\tau_C$ ($\zeta = 6$, $\Delta =  10^{-4}$, $\chi = 0.002$). Colour scale indicates $\left(n_x n_y\right)^2$.  \label{fig:Explicit}}
\end{figure}

This calming effect of polymer on active flow can be reversed by adding direct coupling alongside the kinematic one. Of the two couplings in $f_{QC}$, only the $\chi$ term is sensitive to the relative orientation of tensors $\tsr{C}$ and $\tsr{Q}$; the disruptive case is $\chi>0$ so that these tensors want to be misaligned. \figref{fig:Explicit} shows three novel flow states; for movies see \cite{SI2}. Among these are a shear banded state with interfacial defects (related to those seen in \cite{CristinaDefectPreprint,Thampi2}); coexistence of  `bubbling' active domains and regions with director along the vorticity axis; and states showing periodic modulation of a complex flow pattern on a long time scale set by $\tau_C$, confirming a direct role for polymer viscoelasticity in creating these new states.

\begin{figure}
  \includegraphics[width=0.48\textwidth]{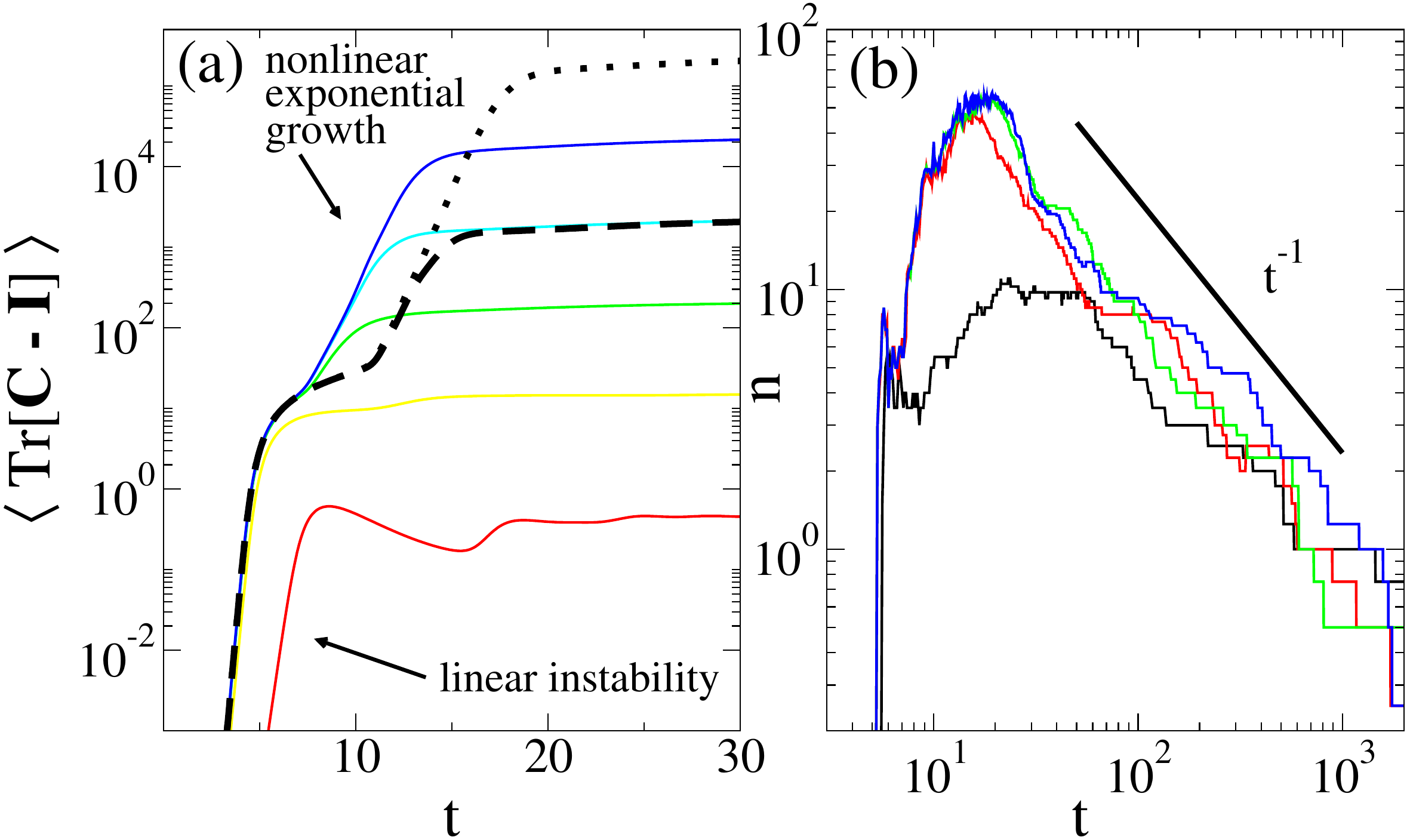}
  \caption{(a) A scalar measure of polymer stress, $\langle\Tr{\tsr{C-I}}\rangle$, against time for $\tau_C = 10^{0}$ (bottom, red) $\to 10^4$ (top, blue) at fixed $\eta_C = 1$. Data for $\tau_C$ infinite, $G_C = 10^{-3}$ (bold dashed), $10^{-5}$ (bold dotted) are also shown.
  (b) Areal defect density $n$ against time for infinite $\tau_C$, with $G_C = 10^{-1}$ (black, bottom) $\to 10^{-7}$ (blue, top); steps arise because $n$ is discrete. In both panels, $\zeta = 3.2$, $\Delta = 8 \times10^{-5}$. }
  \label{fig:Elastomer}
\end{figure}

New and unexpected physics can also arise when this long polymeric time scale becomes effectively infinite, as would describe an active nematic (such as an actomyosin cell extract) within a background of lightly cross-linked elastomer.
We address this limit in two ways: first by increasing $\tau_C$ (holding $\eta_C = 1$), then with $\tau_C$ infinite at small finite $G_C$ (giving infinite $\eta_C$).
The passive limit of this system is a nematic elastomer \cite{Warner}; a full theoretical treatment of the active counterpart will be presented elsewhere \cite{ElastomerStability}. One might expect all of the flow instabilities reported above to be completely absent in what is, after all, a solid material. But this expectation turns out to be misleading. Since $G_C\ll G_Q$, the sample can strongly deform before its small elastic modulus has appreciable effects \cite{PolyGlass}. Accordingly the system should initially show a spontaneous flow instability as though no polymer were present, possibly allowing complex LC textures to form, which then must respond to a growing polymer stress. Numerically (setting $\chi = 0$ for simplicity) we indeed find the onset of spontaneous flow. For $\tau_C \lesssim \tau_Q = 1$ the dynamics is essentially the same as without polymer, and the exponential growth of a shear banding instability is tracked by the polymer stress. We have checked that these observations are stable for small, negative values of $\chi$.

Strikingly, for $\tau_C \gtrsim \tau_Q$, the first phase of exponential growth is followed by a second one (\figref{fig:Elastomer}a), arising because the active turbulent state -- like its passive inertial counterpart -- contains regions of extensional flow where polymers stretch strongly in time.
Although for small $G_C$ large local strains are needed to arrest the spontaneous flow, the time needed to achieve these grows only logarithmically as $G_C \to 0$. For $\tau_C \gg \tau_Q$, rather soon after its initial formation, the turbulent state indeed arrests into a complex but almost frozen defect pattern. Thereafter the defect density decays slowly, roughly as $t^{-1}$ (see \figref{fig:Elastomer}b for $\tau_C \to \infty$ case), which is the classical result for passive nematic coarsening \cite{Bray}.
This process is slow enough that the strain pattern created by the arrested active turbulence might easily be mistaken for a final steady state. Our arrest mechanism, where strong polymer stretching in extensional flow regions creates strong stresses in opposition, may relate closely to the drag reduction effects reported above.

{\it Conclusion:}
To address active viscoelastic matter,
we have created a continuum model combining the theory of active nematics with the well-established Johnson-Segalman (JS) model of polymers. In the passive limit, our model is thermodynamically admissible by design -- a nontrivial achievement since the JS model itself is admissible only by accident. Our model shows that polymers can shift, but not destroy, the generic instability to spontaneous flow shown by active nematics above a critical activity (which still vanishes for large systems). They can also have a strong `bacterial drag reduction' effect, promoting finite throughput in states of active turbulence.

An antagonistic coupling between polymer and nematic orientations produces instead new and complex spontaneous flows, some with oscillation periods set by the polymer relaxation time. Finally, the elastomeric limit of our model reveals, strikingly, that classifying a material as a solid does not {\em a priori} preclude its showing turbulent behavior. Though implausible for inertial turbulence, in the active case this outcome, which arises when $G_C/G_Q \lesssim 0.1$, looks experimentally feasible for subcellular active matter (though probably not swarms of bacteria) within a lightly cross-linked polymer gel.  We hope our work will promote experiments on these and other forms of active viscoelastic matter.

{\it Acknowledgments:} We thank Peter Olmsted for illuminating discussions. E. J. H thanks EPSRC for a Studentship.  M. E. C. thanks EPSRC J007404 and the Royal Society for funding. The research leading to these results has received funding (S. M. F.) from the European Research Council under the European Union's Seventh Framework Programme (FP7/2007-2013) / ERC grant agreement number 279365.  SR acknowledges a JC Bose Fellowship of the DST, India and AM thanks TCIS, TIFR for hospitality.
MCM was supported by the National Science Foundation through awards DMR-1305184 and DGE-1068780 and by the Simons Foundation. The authors thank the KITP at the University of California, Santa Barbara, where they were supported through National Science Foundation Grant NSF PHY11-25925 and the Gordon and Betty Moore Foundation Grant 2919.

{\em Note Added:} After completion of our study a paper appeared addressing similar topics from a somewhat different perspective \cite{addedref}. This treats the spontaneous flow of active particles embedded in a viscoelastic fluid in two dimensions, but unlike our work it (a) omits liquid-crystalline order, and (b) allows for concentration fluctuations. This complementary approach qualitatively confirms some of our findings on bacterial drag reduction.

\clearpage
\pagebreak
\widetext
\begin{center}
  \textbf{Supplemental Material for: \\ \textit{Active Viscoelastic Matter: from Bacterial Drag Reduction to Turbulent Solids}}
\end{center}
\setcounter{equation}{0}
\setcounter{figure}{0}
\setcounter{table}{0}
\setcounter{page}{1}
\makeatletter
\renewcommand{\theequation}{S\arabic{equation}}
\renewcommand{\thefigure}{S\arabic{figure}}
\renewcommand{\bibnumfmt}[1]{[S#1]}
\renewcommand{\citenumfont}[1]{S#1}
\appendix

In this Supplemental Material (SI) we detail the construction of the
equations of our model and describe the linear stability analysis of
the continuum equations. The system under consideration is composed of
orientable particles, endowed with intrinsic dipolar force densities
and dispersed in a polymeric medium. Thus, the theory that we
construct marries polymer physics to active liquid crystal
hydrodynamics \cite{marend2007a}.

\subsection{Configuration Tensors}

For simplicity, we do not keep track of the local concentrations of polymer or active particles and thus are working in the large friction limit in which all components move with the same velocity ${\bf v}$, which is the centre-of-mass velocity. The polymer is modelled by a conformation tensor $\mathbf{C}$ whose departure from isotropy is a measure of local molecular strain.  $\mathbf{C}$ is a second rank tensor that can be defined, depending on the scale of description required, as the dyadic product ${\bf C}=\langle{\bf rr}\rangle$ of the end-to-end vector of an entire polymer, or of a subchain. Comparing some initial configuration (reference space) parameterised by a vector ${\bf R}$, and a deformed configuration (target space) parameterised by a vector ${\bf X}$, gives a deformation $\Lambda_{ij}=\partial X_i/\partial R_j$. A reference space end-to-end vector ${\bf r'}$ then transforms into the target space one as
\begin{equation}
  r'_i=\Lambda_{ij}r_j,
\end{equation}
and an initially isotropic conformation tensor $\langle\vtr{r}\vtr{r}\rangle_{ij} = \delta_{ij}$ transforms to
\begin{equation}
  \langle{\bf r}'{\bf r}'\rangle_{ij}=\Lambda_{ik}\Lambda_{jk},
\end{equation}
which is the left Cauchy-Green tensor.

The orientation of the liquid-crystalline active particles is
described by the traceless symmetric apolar orientational order
parameter $\mathbf{Q}$.

\subsection{Free energy}
The free-energy functional  or effective Hamiltonian of the
system is
\begin{equation}
  \label{free_energy}
  F = \int dV f=\int dV (f_Q + f_C + f_{QC})\;,
\end{equation}
where $f_Q$ is the bare
nematic free-energy density, $f_C$ contains the pure polymeric contribution and
$f_{QC}$ consists of terms that couple the two. For the contribution of the
orientable particles alone we use the standard expression for needle-like
particles (see, e.g., \cite{Lua})
\begin{equation}
  f_Q = G_Q \left[ \frac{(1-\gamma/3)}{2} \mbox{Tr}\mathbf{Q}^2 - \frac{\gamma}{3}
  \mbox{Tr}\mathbf{Q}^3 + \frac{\gamma}{4} (\mbox{Tr}\mathbf{Q}^2)^2 \right] + \frac{K}{2}
  (\nabla_i Q_{jk})^2\;,
  \label{nem_eng}
\end{equation}
and for the polymer part \cite{milnera}
\begin{equation}
  f_C=\frac{G_C}{2}\left(\mbox{Tr}\mathbf{C} -\ln \det \mathbf{C}\right).
  \label{indep}
\end{equation}
Note that Eq. \eqref{indep} describes the dynamics of a collection of
independent chains, with $G_C$ being the osmotic modulus, and can be
augmented if necessary by a standard gradient energy contribution of the form
$(\nabla \mathbf{C})^2$. The possible origin of such a term is discussed below.
To lowest order in gradients and fields, the coupling
between the polymer network and the nematogenic particles is
\begin{equation}
  f_{QC}=\kappa \mbox{Tr}(\mathbf{C}-\mathbf{I})(\mbox{Tr}\mathbf{Q}^2) +
  2\chi\mbox{Tr}\mathbf{CQ}\;,
\end{equation}
constructed to vanish term by term if the network is locally isotropic.

\subsection{The equations of motion}

In this section we write down the equation of motion that govern the
dynamics of the coupled fields $\mathbf{Q}$, $\mathbf{C}$ and $\mathbf{v}$. Each
dynamical equation will contain three
different kinds of couplings: reversible, which cause no relaxation,
irreversible, which in the absence of activity govern the relaxation of the
distribution function to its equilibrium form, and active, encoding the defining
property of the systems of interest here, namely, time-irreversibility at the
level of the individual degrees of freedom.

The dynamics of the apolar
order parameter can be written as
\begin{equation}
  \partial_t Q_{ij}=-v_k\partial_k Q_{ij}+\lambda_{ijkl}\partial_lv_k
  +\Gamma_{ijkl}H_{kl}+\zeta_Q C_{ij}^{ST}\;,
  \label{apolar}
\end{equation}
where superscripts $S$ and $T$ hereafter are used to denote
symmetrization and trace removal, respectively.
The first term on the right-hand side of Eq.~\eqref{apolar} denotes advection by
the flow,
and the second term contains other reversible couplings with the velocity. The
third term is dissipative and couples to the molecular field
\begin{equation}
  \label{QmolfieldH}
  H_{ij} = - \frac{\delta F}{\delta Q_{ij}}^{ST}=
  -G_Q\left[\left(1-\frac{\gamma}{3}\right)\mathbf{Q}-\gamma\mathbf{Q}^2+\gamma
  \mathbf{Q}^3\right]^{ST}_{ij}
  +K\nabla^2Q_{ij}
  -2\kappa
  \rm{Tr}\left(\mathbf{C}-\mathbf{I}\right)Q_{ij}-2\chi\left(\mathbf{C}^T\right)_{
  ij }
  \;.
\end{equation}
We take the kinetic coefficient tensor $\Gamma_{ijkl}$ to be characterized by a
single scalar coefficient:
\begin{equation}
  \label{GammaQ}
  \Gamma_{ijkl}H_{kl} = G_Q \tau^{-1}_Q H_{ij},
\end{equation}
which defines the bare relaxation time $\tau_Q$ for orientational order.
The last term on the right-hand side of Eq.~\eqref{apolar} coupling the nematic
order parameter $\mathbf{Q}$ to the conformational tensor $\mathbf{C}$ is of
active origin.

The conformation tensor obeys the equation of motion
\begin{equation}
  \partial_t C_{ij}=-v_k\partial_k C_{ij}+\lambda^C_{ijkl}\partial_lv_k
  +\Gamma^C_{ijkl}B_{kl}\;.
  \label{polconformation}
\end{equation}
The second and third terms on the right hand side of
Eq.~\eqref{polconformation}, with coefficient $\lambda^C_{ijkl}$ and
$\Gamma^C_{ijkl}$, are
respectively reversible and irreversible couplings and
\begin{equation}
  \label{CmolfieldB}
  B_{ij} = - \frac{\delta F}{\delta C_{ij}}=-\frac{G_C}{2}\left[\delta_{ij}-\left(\mathbf{C}^{-1}\right)_{ij}\right]
  -\kappa\delta_{ij}\rm{Tr}\left(\mathbf{Q}^2\right)-2\chi Q_{ij}
\end{equation}
is the thermodynamic field conjugate to $\mathbf{C}$. We note that although
there is
a term with a coefficient $\zeta_Q$ in Eq.~\eqref{apolar}, there is no
equivalent
active term in Eq.~\eqref{polconformation}. In principle activity allows terms
linear in both $\mathbf{Q}$ and $\mathbf{C}$, in addition to those arising from
couplings in $F[\mathbf{Q},\mathbf{C}]$, in both Eqs.~\eqref{polconformation}
and
\eqref{apolar}. Redefining coefficients within $F$ allows us to absorb all but
one of these. We have chosen, without any loss of generality, to display this
active effect via the $\zeta_Q \mathbf{C}$ term in Eq.~\eqref{apolar}. A term of
a similar nature
could arise in the passive system as well from an allowed free-energy coupling
${\bf C:Q}$. If the dynamics were driven solely by
the free energy, there would then have also been a term proportional to $C_{ij}$
in the $Q_{ij}$ equation, with the same coefficient. The presence of the term
with coefficient $\zeta_Q$ in Eq.~\eqref{apolar}, without a corresponding term
in Eq.~\eqref{polconformation}, breaks the microscopic time-reversal symmetry
of the model. Finally, we have chosen not to retain a dissipative cross-coupling
between $Q_{ij}$ and $C_{ij}$. Such a  cross coupling is allowed by
symmetry and will in general be present, but its inclusion will only result in
inconsequential redefinitions of some phenomenological parameters in our
equations. The explicit expressions for the various dissipative and reactive
coefficients are given in the next subsection.

Ignoring inertia, we take the centre-of-mass velocity to be
determined instantaneously through the dynamics of the other fields.
Total momentum conservation then implies force balance:
\begin{equation}
  \nabla\cdot\boldsymbol{\Sigma}=0\;.
\end{equation}
The total stress tensor
$\boldsymbol{\Sigma}$ is composed of reactive, dissipative and active parts,
\begin{equation}
  \boldsymbol{\Sigma}=\boldsymbol{\Sigma}^{\rm reactive}+\boldsymbol{\Sigma}^{\rm
  diss}+\boldsymbol{\Sigma}_{\rm A}\;.
  \label{stress_tot}
\end{equation}
The
reactive part of the stress is the sum of terms expressed in terms of $C_{ij}$
and $Q_{ij}$,
\begin{equation}
  \boldsymbol{\Sigma}^{\rm reactive}= \boldsymbol{\Sigma}_{\rm Q}+\boldsymbol{\Sigma}_{\rm C}\;,
\end{equation}
with
\begin{eqnarray}
  \label{reactive}
  &&\Sigma_{{\rm Q},ij}=-\Pi_0\delta_{ij} -(\partial_i
    Q_{kl}) \frac{\partial f}{\partial
  \left(\partial_j Q_{kl}\right)} -\lambda_{klij}H_{kl}\;,\\
  &&\Sigma_{{\rm C},ij}=-(\partial_i C_{kl}) \frac{\partial f}{\partial
  \left(\partial_j C_{kl}\right)}-\lambda^C_{klij}{B}_{kl}\;,
\end{eqnarray}
where $\Pi_0$ is the isotropic pressure field, $f$ is the free energy density
[Eq.
\eqref{free_energy}] and the reversible kinetic coefficients $\lambda_{klij}$
and $\lambda^C_{klij}$ will be defined below.
The dissipative part is
\begin{equation}
  \Sigma_{ij}^{\rm diss}=2\eta D_{ij}\;,
\end{equation}
where $D_{ij}$ is the symmetric part of the velocity gradient tensor and $\eta$
the shear viscosity. There are three possible active components of the stress:
\begin{equation}
  \Sigma_{{\rm A},ij}=\Pi_a\delta_{ij}-\zeta Q_{ij}-\zeta_1 C_{ij}\;,
  \label{sigma_active}
\end{equation}
with $\Pi_a$  an active contribution to the (isotropic) pressure. Note that the
active parameter $\zeta_1$ describes polymers that are themselves active, as
opposed to responding to activity through fluid flow, and is set equal to zero
everywhere in the main paper.

Thus, the active terms are $\Pi_a$, $\zeta$ and $\zeta_1$ in the total stress
and $\zeta_Q$ in the $Q_{ij}$ equation.
Every other active term can be subsumed in redefinitions of
free-energy parameters.
Incompressibility renders $\Pi_a$
innocuous. Thus, there remain only three active coefficients which are of
interest, namely, $\zeta$, $\zeta_1$ and $\zeta_Q$. In the main paper we
consider a passive polymer being forced by active nematogenic particles, and
therefore set $\zeta_Q$ and $\zeta_1$ to zero, and only the active stress
coefficient $\zeta$
plays a role in our simulations.

We now take up in turn the free-energy functional and
the coefficients $\lambda_{klij}$ and $\lambda^C_{klij}$ introduced in
Eq. \eqref{reactive}.

\subsection{Reversible couplings}
We expand the reactive kinetic coefficients to first order in the dynamical
variables. We expand $\lambda^C_{ijkl}$ in $\mathbf{C}$ and
$\lambda_{ijkl}$ in $\mathbf{Q}$ only. In general, the kinetic couplings
can depend on both the fields \cite{branda}. However, this too would lead only to
shifts in phenomenological parameters. The same holds for the dissipative
coefficients. To first order in $\mathbf{C}$, we write
\begin{equation}
  \lambda^C_{ijkl}=\frac{a}{2}(\delta_{ik}C_{jl}+\delta_{jk}C_{il}+\delta_{jl}C_{
    ik}+\delta_{il}
    C_{jk})+\frac{1}{2}(\delta_{ik}C_{jl}-\delta_{il}C_{jk}+\delta_{jk}C_{il}
  -\delta_{jl}C_{ik})\;.
\end{equation}
This is not the most general choice for $\lambda^C_{ijkl}$, but we use this to
make contact with the widely used Johnson-Segalman model \cite{JSa}.
For $\lambda_{ijkl}$ we will expand to first order in $\mathbf{Q}$,
\begin{multline}
  \lambda_{ijkl}=
  \frac{\lambda_0}{2}\left(\delta_{ik}\delta_{jl}+\delta_{jk}\delta_{il}-\frac{2}{3}\delta_{ij}
    \delta_{kl}\right)+\frac{1}{2}\left(\delta_{ik}Q_{jl}-\delta_{il}Q_{jk}+\delta_{jk}Q_{il}
    -\delta_{jl}Q_{ik}\right)\\+\frac{\lambda_1}{2}\left(\delta_{ik}Q_{jl}+\delta_{jk}Q_{il}
  +\delta_{jl}Q_{ik}+\delta_{il}Q_{jk}-\frac{4}{3}\delta_{ij}Q_{kl}\right)\;.
\end{multline}
In the simulations we conducted and in the equations we displayed in the main
paper, we further restrict ourselves to the specific case of needle-like
molecules and use the specific form of the flow-coupling coefficients that can be
derived for such particles from microscopics: $\lambda_0=2\xi/3$ and
$\lambda_1=2\xi$ where $\xi$ is a slip parameter. The third flow coupling term
in Eq. (5) of the main paper would require the expansion of $\lambda_{ijkl}$ to
second order in fields. In general, there are many more terms at this order, but
for needle-like particles the coefficient of only one is nonzero (and equal to
$-2\xi$).

\subsection{Irreversible couplings}
The dissipative coefficient in the equation for $\mathbf{Q}$ is expanded only to
to zeroth order in
fields and gradients, thus taking the simple form
$(G_Q\tau_Q)^{-1}\delta_{ik}\delta_{jl}$. This is the value that is displayed in
the main paper.

The dissipative coupling for $\mathbf{C}$ is more complicated since it encodes
the
microstructure of the system. If we assume that the bare conformation tensor
part of
the free energy is the same as for a collection of independent chains, i.e.,
$\int G_C \, \mbox{Tr}(\mathbf{C}-\ln\mathbf{C})$, where $G_C$ is the osmotic
modulus, we
find that a constant, isotropic kinetic coefficient leads to a complicated and unphysical
relaxational form. The conformation-dependent relaxational kinetic
coefficient that Milner \cite{milnera} uses in the case of a polymer gel is
\begin{equation}
  \Gamma^C_{ijkl}=2\tau^{-1}_C\left(\frac{\partial C_{ij}}{\partial
  \Sigma^{el}_{ml}}\right)C_{mk}\;,
\end{equation}
with
\begin{equation}
  \bm\Sigma^{el}=2\mathbf{C}\cdot{\delta F}/{\delta \mathbf{C}}\;.
\end{equation}
This reduces to a kinetic coefficient
\begin{equation}
  \label{Ckincoeff}
  \Gamma^C_{ijkl}=(G_C\tau_C)^{-1}
  (\delta_{ik}C_{jl}+\delta_{il}C_{jk}+\delta_{jl}C_{ik}+\delta_{jk}C_{il})\;,
\end{equation}
if one uses only the bare polymer free-energy, but is much more complicated if
one retains couplings to the apolar order parameter. Nevertheless, we will use
this kinetic coefficient here to construct an active model that reduces to the
J-S
model in the absence of activity. The kinetic coefficient can also be expanded
to higher order in gradients
with terms like $\nabla\nabla\mathbf{C}$, for example. The role of such a term
will be discussed in the next section.

\subsection{Diffusion in JS sector}
In Eq.~(4) of the main paper we have introduced a term
$\ell_C^2\nabla^2\mathbf{C}$ to describe diffusive relaxation of the
conformation tensor. There are at least three mechanisms that can give rise to
this term.
\begin{itemize}
  \item The free-energy \eqref{indep} could in principle be augmented to
    include a term $(\nabla\mathbf{C})^2$, with a coefficient proportional to the
    square of the mesh size. This will then contribute terms of order $\nabla^2
    C_{ij}$
    to $B_{ij}$ in Eq. \eqref{CmolfieldB} if $\Gamma^C_{ijkl}$ in Eq.
    \eqref{polconformation} is evaluated to lowest order in the deformations, i.e.,
    with
    $\mathbf{C} = \mathbf{I}$.
  \item As $\mathbf{C}$ is not a conserved variable, its kinetic
    coefficient $\Gamma^C_{ijkl}$ [see Eq.  \eqref{Ckincoeff}] is nonzero to zeroth
    order
    in gradients. It does, however, receive a contribution to second order in
    gradients
    from the diffusive transport of gel \textit{material}. For a detailed evaluation
    of such a term in the case of the nematic order parameter field see, e.g.,
    \cite{bertina}.
  \item If we were to take the gel concentration into account explicitly, we
    should have to allow for gel currents arising from inhomogeneous polymer
    stresses and hence proportional to $\nabla\cdot\mathbf{C}$. Gradients of such a
    currents would also enter the $\mathbf{C}$ equation of motion, offering one
    more source of terms with two gradients on $\mathbf{C}$.
\end{itemize}
The phenomenological parameter $\ell_C$ in Eq. (4) of the main paper in
principle contains contributions from all of the above mechanisms.

\subsection{Connection to viscoelastic models}
For completeness we also present the bare conformation tensor theory (without
the coupling to $\mathbf{Q}$)  in terms of a dynamical stress, as is often done
in the
polymeric fluid literature. The Johnson-Segalman (JS) model for polymer fluids
\cite{JSa} is one such viscoelastic model. To connect our equations to the JS
model, we rewrite the equation for $\mathbf{C}$ as an equation for the bare
polymer stress,
defined as
\begin{equation}
  \bm\Sigma^0\equiv G_C(\mathbf{C}-\mathbf{I})\;,
  \label{polstress}
\end{equation}
as
\begin{multline}
  \partial_t \mathbf{\Sigma}^0 + \mathbf{v.\nabla \Sigma}^0=  -  2\Gamma_CG{\bm
  \Sigma}^0+{\bm \Sigma}^0{\bm \Omega}-{\bm \Omega}{\bm \Sigma}^0 + 2a
  \left[\mathbf{\Sigma^0 D}\right]^{\text{S}} +2G_Ca\mathbf{D} \\+
  4(G_C\tau_C)^{-1}\left[\kappa \mbox{Tr}[\mathbf{Q}^2]({\bm
        \Sigma}^0+G_C\mathbf{I}) + 2\chi [\mathbf{Q({\bm
  \Sigma}^0+G_C\mathbf{I})}]^S\right] \;.
\end{multline}

Note that by defining polymeric stress as
$[\mathbf{\Sigma}^0]_{ij}=-\lambda^C_{klij}S_{kl}$ we get the same expression as
in
Eq. \eqref{polstress} in the limit of no coupling with the nematic order
parameter
$\mathbf{Q}$. Finally, introducing the renormalized polymer stress
\begin{equation}
  \bm\Sigma^\nu=a\bm\Sigma^0\;,
  \label{polstress_ren}
\end{equation}
we obtain
\begin{multline}
  \partial_t \mathbf{\Sigma}^\nu + \mathbf{v.\nabla \Sigma}^\nu=  -
  2\tau_C^{-1}{\bm \Sigma}^\nu+{\bm \Sigma}^\nu{\bm \Omega}-{\bm \Omega}{\bm
  \Sigma}^\nu + 2a \left[\mathbf{\Sigma^\nu D}\right]^{\text{S}}\\
  +2G_Ca^2\mathbf{D} + 4(G_C\tau_C)^{-1}\left[\kappa \mbox{Tr}[\mathbf{Q}^2]({\bm
        \Sigma}^\nu+a G_C\mathbf{I}) + 2\chi [\mathbf{Q({\bm \Sigma}^\nu+a
  G_C\mathbf{I})}]^S\right] \;.
\end{multline}
In the limit $\mathbf{Q}\to 0$, the above equation reduces to the non-diffusive
J-S model. In order to ensure the correct form for couplings to
other fields such as $\mathbf{Q}$, it seems best, however, to retain the
conformation
tensor rather than expressing it in terms of a polymer stress.

\subsection{Linear stability analysis}

Here we examine the linear stability of an initially non-flowing, homogeneous
base state to $1D$ perturbations in the flow gradient direction, $y$.
The homogeneous base state is described by
\begin{align*}
  \overline{Q}_{\alpha\beta} &= q \left(n_\alpha n_\beta - \delta_{\alpha\beta}
  \right)\;,\\
  \overline{C}_{\alpha\beta} &= \delta_{\alpha\beta}\;,\\
  \overline{\partial_yv_x} &=\overline{ \gdot} = 0\;,
\end{align*}
with $\gdot$ the strain rate. Here $q$ is the magnitude of the order parameter and we choose coordinates such that the director $\uvtr{n} = \left(1, 0, 0\right)$ or $\left(0, 1, 0\right)$, corresponding to a nematic aligned with the flow and flow-gradient directions,
respectively.
To compactify the notation, we introduce a vector $\bm\phi
=\left(\mathbf{Q},\mathbf{C},\gdot\right)$ and perturb the base state by
writing
\begin{align}
  \bm\phi=\overline{\bm\phi}+\delta{\bm\phi}
\end{align}
where
$\overline{\bm\phi}=\left(\overline{\mathbf{Q}},\overline{\mathbf{C}},\overline{
\gdot}\right)$. We write the perturbations
as the sum of Fourier modes
\begin{align}
  \delta\bm\phi= \sum_k \bm\phi^k(t) \textrm{cos}\left(k\pi y / L_y\right)\;,
\end{align}
with Fourier amplitudes $\bm\phi^k=\left(\mathbf{Q}^k,\mathbf{C}^k,\gdot^k\right)$
, and linearize our full set of hydrodynamic equations about the base state to
obtain coupled algebraic equations for the Fourier amplitudes.
Using the Stokes equation,  $\mathbf{\nabla}\cdot\mathbf{\Sigma}=0$, with
$\mathbf{\Sigma}$ the total stress tensor given in Eq. \eqref{stress_tot}, we
can express $\gdot^k$  in terms of ${\tsr{Q}}^k$ and ${\tsr{C}}^k$ as
\begin{align*}
  {\gdot}^k = \frac{-1}{\eta}\left( \delta\tsrt{\Sigma}^k_A+
  \delta\tsrt{\Sigma}^k_Q+ \delta\tsrt{\Sigma}^k_C\right)_{xy},
\end{align*}
where $\delta\mathbf{\Sigma}^k$ denotes the $k$-th Fourier amplitude of the
linearized part of the corresponding contribution to the stress tensor.
Eliminating $\gdot^k$ we finally obtain a linearized  set of algebraic equations
for the six Fourier amplitudes $\vtr{p}^k =  \left({Q}^k_{xx}, {Q}^k_{xy},
{Q}^k_{yy}, {C}^k_{xx}, {C}^k_{xy}, {C}^k_{yy}\right)$ of the form
\begin{equation}
  \partial_t \vtr{p}^k = \tsr{M}^k\cdot\vtr{p}^k\;.
\end{equation}
The eigenvalues of the matrix $\mathbf{M}^k$ yield the dispersion relations
$\omega_k$ of the linear modes of the system as functions of wavevector $k$.
The  real part of such eigenvalues  is the growth rate of the Fourier amplitudes of the perturbations in the hydrodynamic fields. The two non-trivial
eigenvalues are of form $\omega^\pm = -B \pm \sqrt{B^2 - 4 A C} / 2A $, where all quantities are functions of wavevector. The onset of instability corresponds to $\omega^+ = 0$, which simplifies to $A C = 0$. Solving for $\zeta$ yields the critical activity, $\zeta_\textrm{c}$, given in Eq. (7) of the main paper.

\subsection{Throughput definition}
\begin{figure}
  \vspace{30pt}
  \includegraphics[width=0.5\textwidth]{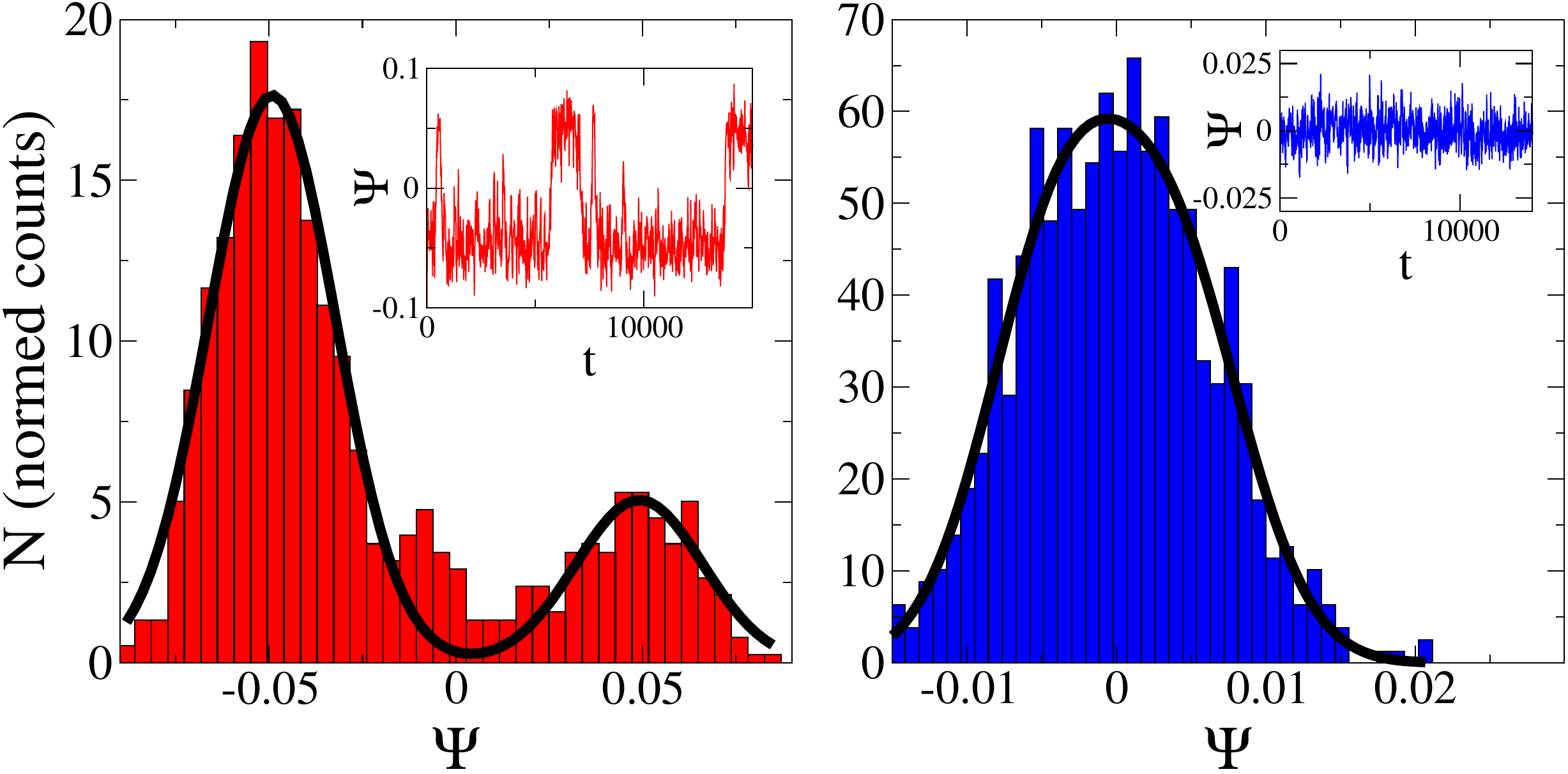}
  \caption{Method for determining throughput. Left panel: a state with net
    throughput, in which the throughput direction switches. The red bins
    show the normalized histogram of $\Psi(t)$, the solid black line is a fit using
    two Gaussians at $\pm \mu_\Psi$. In this example, the positive throughput state
    lasted for shorter simulation time, hence the difference in heights (means and
    standard deviations are the same). Both peaks will tend to the same height in
    the limit $t\to\infty$. Here $\zeta = 5$, $\Delta = 3.2 \times 10^{-4}$, $\tau_C
    = 1$. Right panel: a state with no net throughput for comparison, with $\zeta =
    5$, $\Delta = 10^{-5}$, $\tau_C = 1$. Insets: Examples of throughput-time series
    for each run.
  \label{fig:throughput}}
\end{figure}
We define the throughput as
\begin{equation}
  \Psi(t) = \frac{1}{L_y} \int_0^{L_y} v_x(t) dy = \langle v_x(t) \rangle_y\;.
\end{equation}
As this quantity generally exhibits significant fluctuations in time,
particularly in the chaotic regime, we additionally introduce a  criterion for
`net'  throughput, corresponding to the situation where the mean $\mu_\Psi$
of the throughput histogram exceeds the standard deviation $\sigma_\Psi$.  These
quantities are calculated using a least-squares fit of the throughput histograms
with two Gaussians of width $\sigma_\Psi$,  centered at $\pm \mu_\Psi$ .
Examples of both throughput and non-throughput states are given in shown in
\figref{fig:throughput}.


\begin{thebibliography}{99}
\bibitem{FootIndex} Reversing the order of indices in $\nabla\vtr{v}$ would require sign reversal of $\tsr{\Omega}$ wherever it appears.
\end{thebibliography}
\end{document}